# Studies of YBa$_2$Cu$_3$O$_{6+x}$ degradation and surface conductivity properties by Scanning Spreading Resistance Microscopy

Martin Truchlý[1], Tomáš Plecenik[1], Ondrej Krško[1], Maroš Gregor[1], Leonid Satrapinsky[1], Tomáš Roch[1], Branislav Grančič[1], Marián Mikula[1], Agáta Dujavová[2], Štefan Chromik[2], Peter Kúš[1] and Andrej Plecenik[1]

[1]*Department of Experimental Physics, Faculty of Mathematics, Physics and Informatics, Comenius University, 842 48 Bratislava, Slovakia*
[2]*Institute of Electrical Engineering, Slovak Academy of Sciences, 84104 Bratislava*

**Abstract**

Local surface conductivity properties and surface degradation of c-axis oriented YBa$_2$Cu$_3$O$_{6+x}$ (YBCO) thin films were studied by Scanning Spreading Resistance Microscopy (SSRM). For the surface degradation studies, the YBCO surface was cleaned by ion beam etching and the SSRM surface conductivity map has been subsequently repeatedly measured over several hours in air and pure nitrogen. Average surface conductivity of the scanned area was gradually decreasing over time in both cases, faster in air. This was explained by oxygen out-diffusion in both cases and chemical reactions with water vapor in air. The obtained surface conductivity images also revealed its high inhomogenity on micrometer and nanometer scale with numerous regions of highly enhanced conductivity compared to the surroundings. Furthermore, it has been shown that the size of these conductive regions considerably depends on the applied voltage. We propose that such inhomogeneous surface conductivity is most likely caused by varying thickness of degraded YBCO surface layer as well as varying oxygen concentration (x parameter) within this layer, what was confirmed by scanning Auger electron microscopy (SAM). In our opinion the presented findings might be important for analysis of current-voltage and differential characteristics measured on classical planar junctions on YBCO as well as other perovskites.

**1. Introduction**

A lot of effort has been done over last approximately 25 years to understand conductivity properties of complex transition metal oxides, particularly high temperature superconductors (HTS) e.g. YBa$_2$Cu$_3$O$_{6+x}$ (YBCO) or Bi$_m$Sr$_2$Ca$_{n-1}$Cu$_n$O$_{2n+m+2}$ (BSCCO), in both superconducting and normal state due to their multifunctional nature and many potential applications. It is known that among other factors, the oxygen content *x* is a key parameter determining the conductivity properties of these compounds [1]. For YBCO, by varying the parameter *x* between 1 and 0 the conductivity properties are continuously changing from metallic through semiconducting to insulating (above the critical temperature T$_c$, which also depends on *x*). Below the Néel temperature, an insulating phase with antiferromagnetic ordering is observed for small values of x [1].

Relatively low activation energy for the oxygen diffusion in the YBCO, which has been shown by various experiments to be in range from ~0.5 to ~1.3 eV within the material and about ~1.7 eV for out-diffusion from the surface [2-5], leads to leakage of the oxygen out of the YBCO surface and spontaneous creation of a degraded insulating near-surface layer with decreased amount of oxygen content [5-7]. This is a big challenge in many applications where well defined and sharp interfaces between HTS and electrodes are required. Several experiments studying the degradation processes in terms of gradual increasing of an YBCO/metal junction resistance due to oxygen loss has been done. Most of these experiments showed increasing of the junction resistance over time until a saturation level is reached [5]. Such degradation has not been observed only at low temperatures where the oxygen diffusion and out-diffusion is significantly suppressed [8]. However, it was not possible to study the YBCO surface degradation on air by such experiments due to necessity of placing a metallic electrode on the surface. Moreover, such methods are capable of measuring only an average current through the electrode area and no lateral distribution of the measured current can be observed. Measurements of the current-voltage (I-V) and differential (dI/dV-V) characteristics on such normal metal – superconductor (NS) and normal metal – insulator – superconductor (NIS) junctions thus assumed homogeneous current density distribution over the junction area. Indeed, some insight into local surface properties of YBCO and other perovskites has been provided by numerous scanning tunneling microscopy (STM) measurements [9-12]. However, as the STM method is based on measurement of the tunnel barrier resistance, it is very hard to reliably distinguish between influence of topography and thickness and strength of the insulating barrier formed at the YBCO surface. As the YBCO surface degrades in air [13] and even in ultra-high vacuum (UHV) conditions [9] due to oxygen loss, interpretation of STM images and scanning tunneling spectroscopy (STS) spectra is often difficult due to unknown local thickness and conductivity of the degraded layer which can partly substitute the vacuum tunneling barrier. Authors thus usually do not claim vacuum tunneling in room temperature measurements, report bad tunnel contacts [11] or even admin direct contact of the STM tip with the surface in some cases [9,12,14]. Reproducible atomic-resolution STM studies of clean non-degraded YBCO surfaces with tunneling through well-characterized vacuum barrier were however reported at single crystals cleaved in-situ at 20 K and measured at low temperatures [10].

Here we present surface degradation and local surface conductivity studies of c-axis oriented YBCO thin films by the Scanning Spreading Resistance Microscopy (SSRM) method, which allows us to study the surface conductivity properties in air or other atmosphere (the surface does not has to be covered by an electrode) and with high lateral resolution of the current density distribution. Unlike in STM where topography (i.e. tip-sample separation) and possible changes in surface conductivity are mixed together in the measured tunneling resistance signal, in SSRM the topography and conductivity are measured independently, what allows reliable surface conductivity mapping even on degraded surfaces. In this work we have shown by the SSRM that at constant tip-sample voltage the current density distribution over the surface of our YBCO samples is highly inhomogeneous with numerous regions of highly enhanced conductivity compared to the surroundings. Moreover, size of these regions was shown to considerably depend on the applied voltage. After removing the degraded surface layer by ion beam etching, it has been also shown that the average surface conductivity gradually decreases in time in both air and pure nitrogen atmosphere due to oxygen loss as expected. We propose that such inhomogenity of the surface conductivity should be taken into account when analyzing I-V and dI/dV-V characteristics measured on typical μm-scale planar junctions on YBCO and other perovskites and that it can also explain some peculiarities attributed to HTS, e.g. smeared gap-like structure.

## 2. Experimental

High-quality *c*-axis oriented YBCO thin films with thickness of about 200 nm were prepared by dc magnetron sputtering from a stoichiometric ceramic target on single-crystalline $LaAlO_3$ substrates. Sputtering process was carried out in an oxygen atmosphere at a pressure of 240 Pa, a magnetron power of 4.5 W/cm$^2$ and the substrate temperature of 815 °C. The samples were subsequently annealed in oxygen at a pressure of $10^4$ Pa at 500 °C during 30 min. After that, cooling to room temperature at a rate of 15 °C/min was performed. Critical temperature $T_c$ of the prepared films has been checked to be about 85 K. The c-axis orientation of the films has been verified by x-ray diffraction (PANalytical X'Pert PRO diffractometer). The films were further examined in cross-section by field-emission scanning electron microscope (SEM) available in TESCAN Lyra FIB-SEM system. For the degradation studies, Ar ion beam etching (ion gun Platar Klan 53M, ion energy 500 eV) was used to achieve minimal initial degradation of the YBCO surface.

For the topography and surface conductivity measurements, scanning probe microscopes (SPM) NT-MDT Solver P47 Pro and NT-MDT NTEGRA Aura were used. The surface topography has been measured in semicontact Atomic Force Microscopy (AFM) mode and the surface conductivity mapping has been done in Scanning Spreading Resistance Microscopy (SSRM) mode, which also simultaneously provided contact AFM topographic images. Conductive PtIr-coated silicon AFM tips have been used. In the SSRM method, a bias voltage is applied between the STM tip and the sample during contact AFM topography imaging and the resulting current flowing between the tip and the sample is measured. At a given voltage, this current is given by the total circuit resistance $R_{total}$ composed of the tip resistance $R_t$, the sample-tip contact resistance $R_c$ and the sample resistance $R_s$ (considering wire and other resistances negligible): $R_{total} = R_t + R_c + R_s$. As the circuit cross-section is smallest at the tip-sample contact, the $R_t$ and $R_s$ resistances can be usually considered constant and negligible compared to $R_c$ for conductive tips and reasonable sample resistance. Obtained SSRM images thus reflect local conductivity of a near-surface (several nm) layer with lateral resolution given by the tip-sample contact area. During the SSRM measurements the YBCO thin films were biased by the internal SPM voltage source and the SPM tip was grounded. For the surface degradation studies, the bias voltage was set to 1.5 V, while for the investigation of the voltage dependence of a one particular high-conductivity region, the bias voltage varied in the range from 0 to +10 V.

To correlate the surface conductivity with the amount of oxygen (x parameter) within the YBCO surface cells, Auger electron spectroscopy (AES) and scanning Auger electron microscopy (SAM) available in the Omicron MULTIPROBE UHV system with SPHERA electron spectrometer have been used. This system is also equipped by ion gun for in-situ ion beam etching.

## 3. Results and discussion

Typical AFM topography and corresponding SSRM images (1.5 V bias voltage) of the YBCO films stored on air for several days are shown on Fig. 1. The SSRM image clearly shows that the YBCO surface conductivity is very low due to highly degraded (the *x* parameter is close to zero) near-surface layer and some current flows only through areas of the highest surface irregularities.

After the degraded layer has been removed by the ion beam etching, the average surface conductivity considerably increased although the current distribution over the sample surface was highly inhomogeneous (Fig. 2a). For the degradation studies, the samples were stored in air or pure nitrogen atmosphere and selected area of 5x5 μm has been repeatedly scanned by the SSRM over several hours after the ion beam etching and the current / conductivity map has been recorded for each scan (Fig. 2). The SSRM images clearly show that the average conductivity of the cleaned YBCO surface gradually decreases in time in both air and pure nitrogen atmosphere. The time dependence of the average current density calculated from the SSRM images (Fig. 3) shows its approximately linear decrease in time over first 8 hours after the initial degraded layer was removed by the ion beam etching. The degradation was slightly slower in nitrogen atmosphere due to lack of water vapor which causes additional degradation of the YBCO surface due to various chemical reactions [9]. However, the surface still undergoes degradation due to oxygen out-diffusion which, as was pointed above, is suppressed only at low temperatures.

The observed decrease of the surface conductivity over time is in agreement also with works done on YBCO/metal contacts [5]. Effects shown in Fig. 2 and 3 show that the shape of I-V and dI/dV-V characteristics measured on typical μm-sized junctions can change in time, particularly within the first several hours after the preparation of the junction.

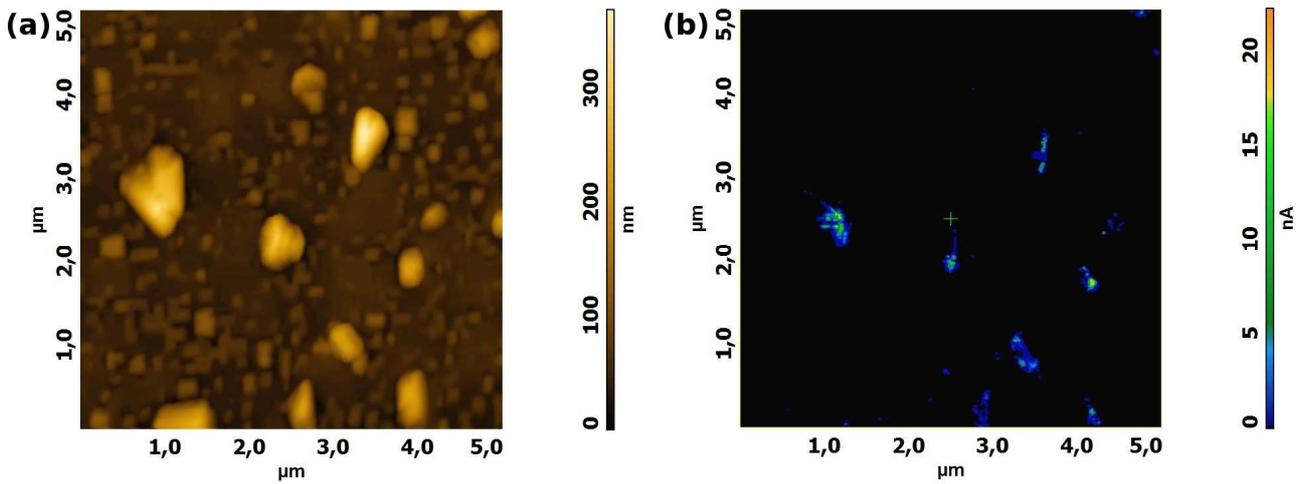

Fig. 1. (Color online) Typical (a) AFM topography and (b) SSRM image (bias voltage 1.5 V) of the YBCO surface stored in air for several days before the ion beam etching.

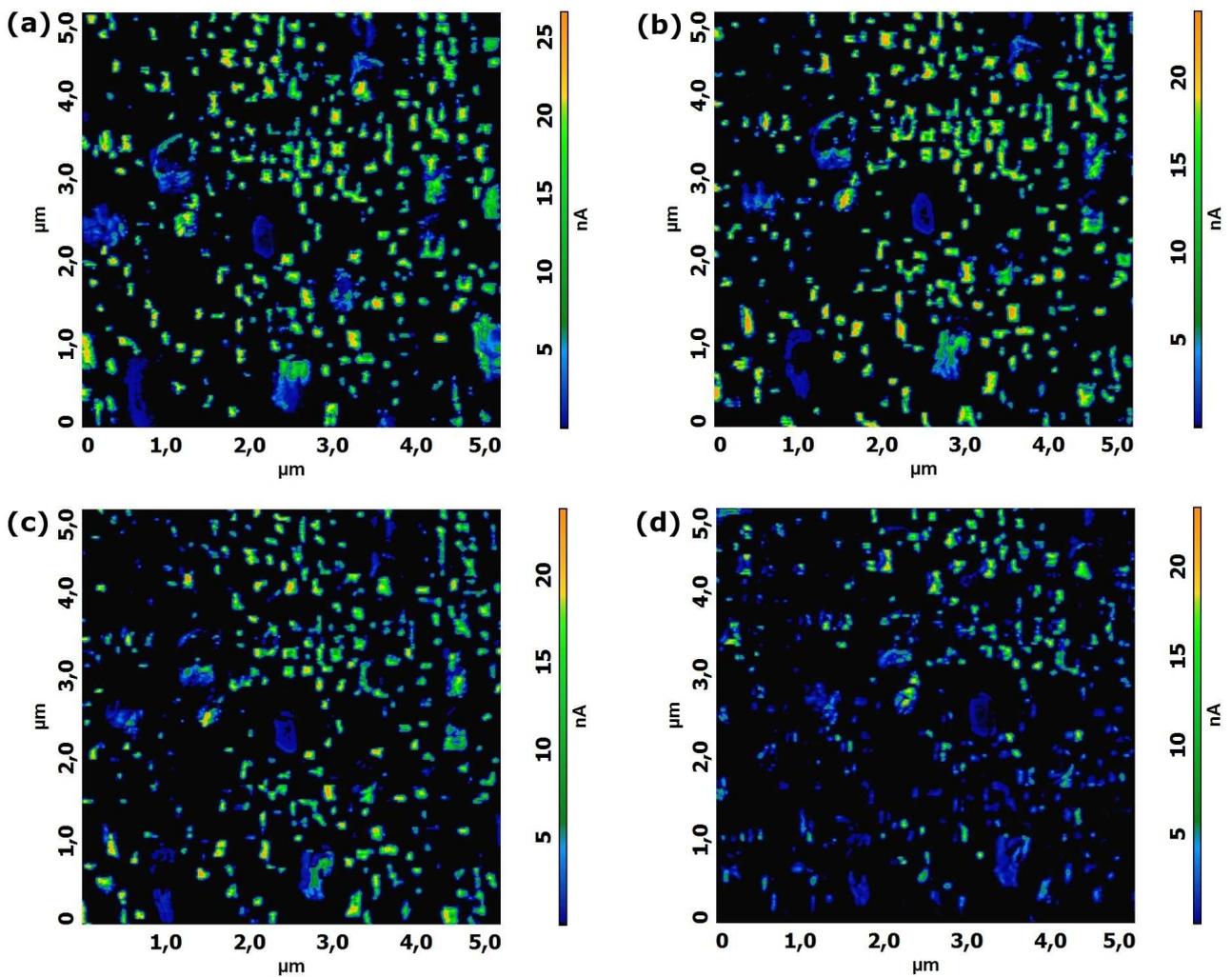

Fig. 2. (Color online) SSRM images (bias voltage 1.5 V) of the YBCO surface after (a) 1.5 hours (b) 4 hours (c) 7 hours (d) 9 hours in air after the sample was etched in ion beam. Bright areas indicate regions with high conductivity.

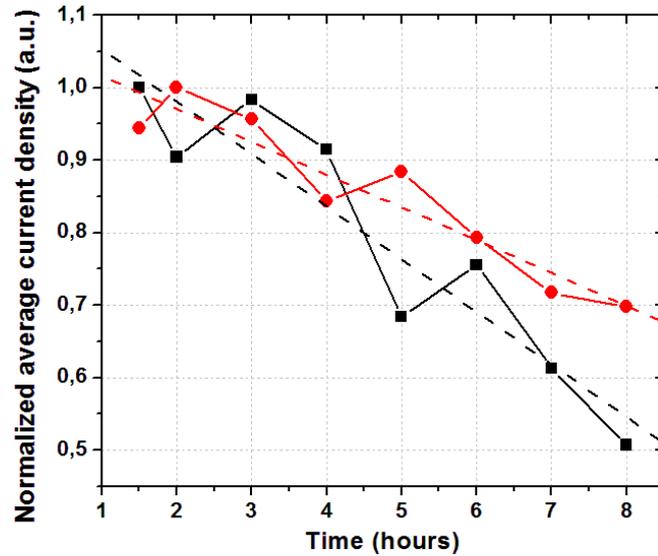

*Fig. 3. (Color online) Average SSRM current (normalized to maximum value) measured on YBCO surfaces stored in air (black squares) and pure nitrogen (red circles) after the ion beam etching calculated from the 5x5 μm SSRM images as a function of time. The linear fits (dashed lines) are a guide for an eye.*

Subsequently, we have focused on a selected surface region with one of the areas with higher conductivity, which generally had an irregular shape. In general, their conductivity, shape and size considerably varied with applied voltage (Fig. 4). At bias voltage of +1V, the high-conductivity area was practically not visible on the SSRM image and its conductivity appeared to be the same as the surroundings (Fig. 4a). With increasing voltage to +2V, its conductivity became higher compared to surroundings and it became visible (Fig. 4b). With further increasing of the applied voltage, both the conductivity and size of this high-conductivity area were gradually increasing (Fig. 4c-f). Conversely, conductivity of the surrounding surface remained very low and practically immeasurable.

Existence and behavior of such areas can be explained by varying thickness of the degraded insulating surface layer. These variations are most likely given by the polycrystalline nature of the film, where near-optimally doped grains of various shape and size may be present under the insulating layer. Depending on their size, shape and placement, the thickness of the degraded insulating barrier can vary from zero (direct contact) through very thin (~ 1 nm) where the transport properties are given by a tunneling current, up to relatively thick (several nanometers) completely insulating barrier. As the oxygen diffusion is expected to be faster along the crystalline boundaries, we propose that the polycrystalline nature of the film most likely does not lead only to variable thickness of the degraded layer, but also to inhomogeneous oxygen distribution (x-parameter) in the CuO chains of the YBCO unit cells within the surface degraded layer itself. This, according to the YBCO T-x phase diagram may also have significant effect on the surface conductivity. To support these propositions the YBCO films were examined in cross-section by SEM to confirm their polycrystalline nature (Fig. 5a). Furthermore the surface of the films was analyzed by AES and SAM after in-situ cleaning by ion-beam etching (800 eV ion energy) in UHV chamber. The AES spectra (Fig. 5b) did not show any surface contamination (e.g. carbon) which might have led to variation of oxygen concentration on the surface. However, inhomogeneous oxygen distribution on micrometer scale has been confirmed by SAM by mapping intensity of the oxygen Auger peak over the sample surface (Fig. 5c).

Such highly conductivity-inhomogeneous regions can considerably influence properties of Josephson junctions prepared by planar technology as well as measurements of I-V characteristics and density of states (DOS) by tunneling spectroscopy. Particularly on the junctions prepared by deposition of normal metal as an upper electrode directly on HTS, the parallel conducting channels can be created within one junction with large area (μm-scale). Thus in one junction can co-exist SIN structures with tunneling current $I_T$, SN structures exhibiting Andreev reflection and current $I_A$,

SS'N with current $I_D$, SN'N, etc., where S is non degraded HTS, S' is HTS with depressed superconductivity, N' is degraded HTS in normal state, I is HTS in the state of insulator or antiferromagnetic insulator (see YBCO T-x phase diagram in Ref. 1) and N is normal metal. As the total current is expressed in the form $I = \alpha I_T + \beta I_A + \gamma I_D + ...$, the shape of the final I-V and dI/dV-V curves depends on the weight coefficients only and one can restore differential characteristics with peculiarities attributed to HTS, i.e. smeared gap-like structure etc.

It should be noted that on the areas with high surface roughness some inhomogenities can be also caused by contact of the SPM probe with the ab-plane (which has about 30 times higher conductivity [1]) of the YBCO even on well oriented crystals. Moreover, at the higher voltages the measurements can be influenced also by relatively strong electrostatic force affecting the AFM tip as well as electrostatic field induced diffusion of oxygen ions within the YBCO.

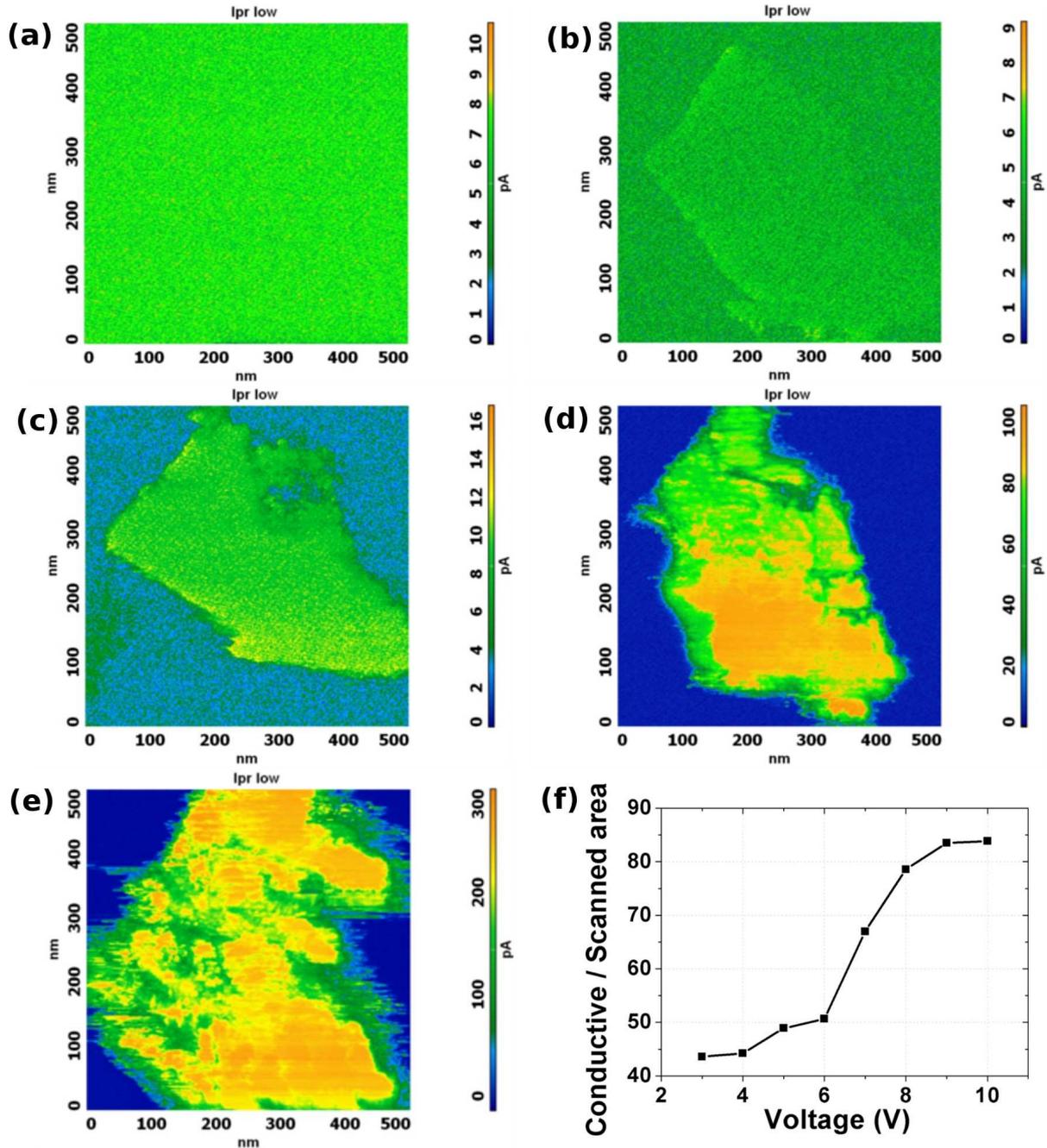

*Fig. 4. (Color online) SSRM images of a selected region with one of the high-conductivity areas on YBCO surface at voltages (a) 1 V; (b) 2 V; (c) 4 V; (d) 6 V; (e) 8 V. (f) Size of the high-conductivity area normalized to the scanned area as a function of the applied voltage.*

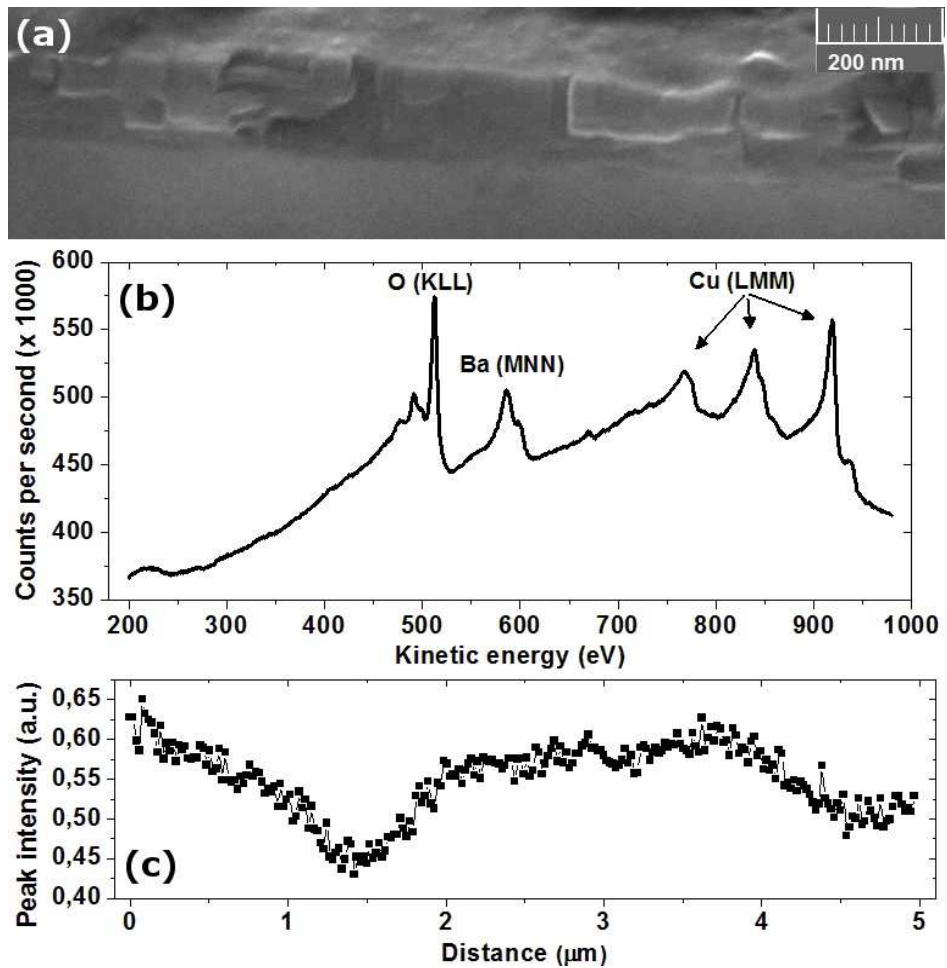

*Fig. 5. (a) Cross-section SEM micrograph of the YBCO film. (b) Part of the Auger spectra of the YBCO surface after in-situ surface cleaning by ion beam etching. (c) SAM oxygen peak intensity scan over 5 μm line normalized as (peak – background) / (peak + background) to minimize influence of the topography.*

## 4. Conclusions

To conclude, a lateral distribution of the surface conductivity and the time evolution of the surface degradation of c-axis oriented YBCO thin films have been studied by the SSRM method. It has been shown that the surface conductivity of our YBCO films is highly inhomogeneous on micrometer and sub-micrometer scale, with areas of considerably higher conductivity compared to the surroundings. After removing the degraded layer, the surface conductivity considerably increased, but then gradually degraded again within several hours due to oxygen out-diffusion in both air and pure nitrogen atmosphere. Furthermore it has been shown that the size of the high-conductivity areas is considerably increasing with increasing of the applied voltage. We propose that such inhomogeneous surface conductivity is caused by a degraded oxygen-depleted surface layer of varying thickness and inhomogeneous oxygen concentration distribution within the CuO chains of the YBCO unit cells on the surface. This was also confirmed by scanning Auger electron microscopy. In our opinion, such inhomogenity of the surface conductivity should be taken into account when analyzing I-V and dI/dV-V characteristics measured on typical μm-scale planar junctions on YBCO and other perovskites. We further propose that it can also explain some peculiarities attributed to HTS such as smeared gap-like structure. Moreover, it can be also a limiting factor for nano-scale applications.

**Acknowledgements**

This project has been funded with support from the European Commission Grant No. EC NMP3-SL-2011-283141 — IRON SEA. This publication reflects the views only of the authors, and the Commission cannot be held responsible for any use which may be made of the information contained therein. This work was also supported by the Slovak Research and Development Agency under Contract No. LPP-0176-09 and is also the result of the project implementation: 26220220004 supported by the Research & Development Operational Programme funded by the ERDF.